\documentclass[%
 aip,
 amsmath,amssymb,
 reprint,%
]{revtex4-2}

\usepackage{amsthm}
\usepackage{amsfonts}
\usepackage{amssymb}
\usepackage{graphicx}
\usepackage{dcolumn}
\usepackage{bm}

\usepackage[utf8]{inputenc}
\usepackage[T1]{fontenc}
\usepackage{mathptmx}
\usepackage{xcolor}

\usepackage[unicode=true,
 bookmarks=true,bookmarksnumbered=false,bookmarksopen=false,
 breaklinks=false,pdfborder={0 0 1},backref=false,colorlinks=true]
 {hyperref}
\hypersetup{
 linkcolor=magenta, urlcolor=blue, citecolor=blue, pdfstartview={FitH}, unicode=true}

\def\dd{{\mathrm{d}}}
\def\erfc{{\mathrm{erfc}}}

\begin{document}

\preprint{AIP/123-QED}

\title[Directed percolation and numerical stability of simulations of digital memcomputing machines]{Directed percolation and numerical stability of simulations of digital memcomputing machines}

\author{Yuan-Hang Zhang}
\email{yuz092@ucsd.edu}
\affiliation{Department of Physics, University of California, San Diego, CA 92093, USA}
\author{Massimiliano Di Ventra}
\email{diventra@physics.ucsd.edu}
\affiliation{Department of Physics, University of California, San Diego, CA 92093, USA}

\date{\today}

\begin{abstract}
Digital memcomputing machines (DMMs) are a novel, non-Turing class of machines designed to solve combinatorial optimization problems. They can be physically realized with continuous-time, non-quantum dynamical systems with memory ({\it time non-locality}), whose ordinary differential equations (ODEs) can be numerically integrated on modern computers. Solutions of many hard problems have been reported by numerically integrating the ODEs of DMMs, showing substantial advantages over state-of-the-art solvers. To investigate the reasons behind the robustness and effectiveness of this method, we employ three explicit integration schemes (forward Euler, trapezoid and Runge-Kutta 4th order) with a constant time step, to solve 3-SAT instances with planted solutions. We show that, {\it (i)} even if most of the trajectories in the phase space are destroyed by numerical noise, the solution can still be achieved; {\it (ii)} the forward Euler method, although having the largest numerical error, solves the instances in the least amount of function evaluations; and {\it (iii)} when increasing the integration time step, the system undergoes a “solvable-unsolvable transition” at a critical threshold, which needs to decay at most as a {\it power law} with the problem size, to control the numerical errors. To explain these results, we model the dynamical behavior of DMMs as directed percolation of the state trajectory in the phase space in the presence of noise. This viewpoint clarifies the reasons behind their numerical robustness and provides an analytical understanding of the solvable-unsolvable transition. These results land further support to the usefulness of DMMs in the solution of hard combinatorial optimization problems.	
\end{abstract}

\maketitle

\begin{quotation}
Certain classical dynamical systems with memory, called digital memcomputing machines, have been suggested to solve a wide variety of combinatorial optimization problems~\cite{DMMperspective}. 
In particular, simulations of their ordinary differential equations (ODEs) on traditional computers, using explicit methods of integration, have shown substantial advantages over state-of-the-art algorithms. These results are remarkable considering the fact that the ODEs of these dynamical systems are stiff, hence should become highly unstable if integrated with explicit methods. In 
this work, by drawing a connection between directed percolation and state dynamics in phase space, we provide both numerical and analytical reasons why explicit methods of integration for these ODEs are robust against the unavoidable numerical errors. 
\end{quotation}
\section{Introduction}
\label{sec:intro}
MemComputing is a recently proposed (non-Turing) computing paradigm in which {\it time non-locality} (memory) accomplishes both tasks of information 
processing and storage~\cite{diventra13a}. Its {\it digital} (hence scalable) version (digital memcomputing machines or DMMs) has been introduced specifically to solve combinatorial 
optimization problems~\cite{DMM2,DMMperspective}. 

The basic idea behind DMMs is that, instead of solving a combinatorial optimization problem in the traditional algorithmic way, one maps it into a specifically designed dynamical system with memory, so that the only equilibrium points of that system represent the solutions of the given problem. This is accomplished by writing the problem in Boolean (or algebraic) form, and then replacing the traditional (unidirectional) gates with {\it self-organizing gates}, that can satisfy their logical (or algebraic) relation {\it irrespective} 
of the direction of the information flow, whether from the traditional input or the traditional output: they are {\it terminal-agnostic gates}.~\cite{DMM2,DMMperspective,ILP,Branching}. In addition, the resulting dynamical 
systems can be designed so that no periodic orbits or chaos occur during dynamics~\cite{no-chaosa,seanpaper}. 

DMMs then map a {\it finite} string of symbols into a {\it finite} string of symbols, but operate in {\it continuous time}, {hence they are distinct from Turing machines that operate in discrete time. (Note that universal memcomputing machines have been shown to be Turing-complete, but not Turing-equivalent~\cite{UMM}.)} As a consequence, they 
seem ideally suited for a hardware implementation. In fact, they can be realized in practice with non-linear, {\it non-quantum} dynamical systems with memory, such as electrical circuits implemented with conventional complementary metal–oxide–semiconductor technology~\cite{DMM2}. 

On the other hand, DMMs, being classical dynamical systems, are such that their state trajectory   
belongs to a topological manifold, known as {\it phase space}, whose dimension, $D$, scales {\it linearly} with the number of degrees of freedom~\cite{Goldreich}. 
In addition, their state dynamics are described by {\it ordinary differential equations} (ODEs)~\cite{DMM2,DMMperspective}. One can then attempt to {\it numerically} integrate these ODEs on our traditional computers, using any 
integration scheme~\cite{Sauer}. However, naively, one would expect that the computational overhead of integrating these ODEs, coupled with  
large numerical errors, would require an unreasonable amount of CPU time as the problem size grew, hence limiting the realization of DMMs (like quantum computers) to only hardware. 

To make things worse, the ODEs of DMMs are {\it stiff}~\cite{Sauer}, meaning that 
they have several, quite distinct, intrinsic time scales. This is because the memory variables of these machines have a much slower dynamics than the degrees of freedom representing the logical symbols (variables) of the problem~\cite{DMM2,DMMperspective}. Stiff ODEs are notorious for requiring 
{\it implicit} methods of integration, which, in turn, require costly root-finding algorithms (such as the Newton's method), thus making the numerical simulations very challenging~\cite{Sauer}. 

This leads one to expect poor performance when using {\it explicit} methods, 
such as the simplest {\it forward Euler} method, on the ODEs of the DMMs. Instead, 
several results using the forward Euler integration scheme on DMMs have shown that these machines still find solutions to a variety of combinatorial optimization problems,
including, Boolean satisfiability (SAT)~\cite{seanpaper}, maximum satisfiability~\cite{exponential2017speedup,stress-test}, spin glasses~\cite{spinglass},
machine learning~\cite{AcceleratingDL,UnsupervisedL}, and integer linear programming~\cite{ILP}.

For instance, in Ref.~\onlinecite{seanpaper}, the numerical simulations of the DMMs, using forward Euler with an adaptive time step, substantially outperform traditional algorithms~\cite{selman1993walksat, mezard2002sid,yalsat}, in the solution of 3-SAT instances with planted solutions. The results show a {\it power-law} scaling in the number of integration steps as a function of problem size, avoiding the exponential scaling seen in the performance of traditional algorithms on the same instance classes. Furthermore, in Ref.~\onlinecite{seanpaper}, it was found that the average size of the adaptive time step decayed as a {\it power-law} as the problem size grew. 

It is then natural to investigate how the power-law scaling of integration steps varies with the numerical integration scheme employed to simulate the DMMs. In particular, by employing a constant integration time step, $\Delta t$, we would like to investigate if the requisite time step to solve instances and control the numerical errors, continues to decay with a power-law, or will require exponential decay as the problem grows.

In this paper, we perform the above investigations while attempting to solve some of the planted-solution 3-SAT instances used as benchmarks in Ref.~\onlinecite{seanpaper}.
We will implement three {\it explicit} integration methods (forward Euler, trapezoid, and Runge-Kutta 4th order) while numerically simulating the DMM dynamics to investigate the effect on scalability of the number of integration steps versus the number of variables at a given clause-to-variable ratio.

Our numerical simulations indicate that, regardless of the explicit integration method used, the ODEs of DMM are robust against the numerical errors caused by the discretization of time.
The robustness of the simulations is due to 
the {\it instantonic dynamics} of DMMs~\cite{topo,DMtopo}, 
which connect critical points in the phase space with increasing stability. {Instantons in dissipative
systems are specific heteroclinic orbits connecting two distinct
(in index) critical points.} Both the critical points and the instantons connecting them 
are of topological character~\cite{Solitons,Coleman}. Therefore, if the integration method preserves critical points (in number and index) then the solution search is ``topologically protected'' 
against reasonable numerical noise~\footnote{Incidentally, this is also the reason why the {\it hardware} implementation of DMMs would be robust against 
	reasonable {\it physical} noise~\cite{topo,DMtopo}.}. 

For each integration method, we find that when increasing the integration time step $\Delta t$, the system undergoes a ``solvable-unsolvable transition''~\footnote{Note that all instances can be solved, as they have planted solutions.} at a critical $\Delta t_c$, which 
decays at most as a {\it a power law} with the problem size, avoiding the undesirable exponential decay that severely limits numerical simulations. 
We also find that, even though higher-order integration schemes are more favorable in most scientific problems, the first-order forward Euler method works the best for the dynamics of DMMs, providing best scalability in terms of function evaluations vs. the size of the problem. We attribute this to the fact that the forward Euler method preserves critical points, irrespective of the size of the integration time step, while higher order methods may introduce ``ghost critical points" in the system \cite{cartwright1992dynamics}, disrupting the instantonic dynamics of DMMs. 

Finally, we provide a physical understanding of these results, by showing that, in the presence of noise, the dynamical behavior of DMMs can be modeled as a {\it directed percolation} of the state trajectory in the phase space, with the inverse of the time step $\Delta t$ playing the role of percolation probability. We then analytically show that, with increasing percolation probability, the system undergoes a {\it permeable-absorbing phase transition}, which resembles the ``solvable-unsolvable transition'' in DMMs. All together, these results clarify the reasons behind the numerical robustness of the simulations of DMMs, and further reinforce the notion that these dynamical systems with memory are a useful tool for the solution of hard combinatorial optimization problems, not just in their hardware implementation but also in their 
numerical simulation. 

This paper is organized as follows. In Sec.~\ref{sec:DMM} we review the DMMs used in Ref.~\onlinecite{seanpaper} to find satisfying assignments to 3-SAT instances. (Since our present paper is not a benchmark paper, the interested reader is directed to Ref.~\onlinecite{seanpaper}
for a comparison of the DMM approach to traditional algorithms for 3-SAT.) 
In Sec.~\ref{sec:scalability} we compare the results on the scalability of the number of steps to reach a solution for three explicit methods: forward Euler, 
trapezoid, and Runge-Kutta 4th order. For all three methods, our simulations show that increasing $\Delta t$ (thereby increasing numerical error) induces a sort of ``solvable-unsolvable phase transition,'' that becomes sharper with increasing size of the problem. However, we also show that the ``critical'' time step, $\Delta t_c$, decreases as a {\it power law} as the size of the problem increases for a given clause-to-variable ratio. In Sec.~\ref{sec:DP}, we model this transition 
as a {\it directed percolation} of the state trajectory in phase space, with $\Delta t$ playing the role of the inverse of the percolation probability. 
We conclude in Sec.~\ref{sec:conclusion} with thoughts about future work. 

\section{Solving 3-SAT with DMMs}
\label{sec:DMM}

In the remainder of this paper, we will focus on solving satisfiable instances of the 3-SAT problem \cite{barthel2002}, 
which is defined over $N$ Boolean variables $\{y_i=0, 1\}$ constrained by $M$ clauses. 
Each clause consists of 3 (possibly negated) Boolean variables connected by logical OR operations, and 
an instance is solved
when an assignment of $\{y_i\}$ is found such that all $M$ clauses evaluate to TRUE (satisfiable).

For completeness, we briefly review the dynamical system with memory representing our DMM.
The reader is directed to Ref.~\onlinecite{seanpaper} for a more detailed description and its mathematical properties.

To find a satisfying assignment to a 3-SAT instance with a DMM, we transform it into a Boolean circuit,  where the Boolean variables $y_i$ are transformed into continuous variables and represented with terminal voltages $v_i\in [-1, 1]$ (in arbitrary units), where $v_i>0$ corresponds to $y_i=1$, and $v_i<0$ corresponds to $y_i=0$. We use $(l_{m, i}\!\vee l_{m, j}\!\vee l_{m, k})$ to represent the $m$-th clause, where $l_{m, i}=y_i$ or $\bar{y}_i$ depending on whether $y_i$ is negated in this clause. Each Boolean clause is also transformed into a continuous constraint function:
\begin{equation}
    C_m(v_i, v_j, v_k) = \frac{1}{2}\min\left(1-q_{m, i}v_i, 1-q_{m, j}v_j, 1-q_{m, k}v_k\right),
\end{equation}
where $q_{m, i}=1$ if $l_{m_i}=y_i$ and $q_{m, i}=-1$ if $l_{m_i}=\bar{y}_i$. We can verify that the $m$-th clause evaluates to true if and only if 
$C_m<0.5$~\cite{seanpaper}. 

The idea of DMMs is to propagate information in the Boolean circuits `in reverse' by using self-organizing logic gates (SOLGs) \cite{DMM2, DMMperspective} 
(see~\cite{ILP} for an application of {\it algebraic} ones). 

SOLGs are designed to work bidirectionally, a property referred to as ``terminal agnosticism''~\cite{Branching}. The terminal that is traditionally an output can now receive signals like an input terminal, and the traditional input terminals will self-organize to enforce the Boolean logic of the gate. 
This extra feature requires additional (memory) degrees of freedom within each SOLG. We introduce two additional memory variables per gate: ``short-term'' memory $x_{s, m}$ and ``long-term'' memory $x_{l, m}$. For a 3-SAT instance, the dynamics of our self-organizing logic circuit (SOLC) are governed by Eq. \eqref{eq:DMM} \cite{seanpaper}: 

\begin{equation}
\begin{aligned}
&\dot{v}_{i}=\sum_{m} x_{l, m} x_{s, m} G_{m, i}+\left(1+\zeta x_{l, m}\right)\left(1-x_{s, m}\right) R_{m, i}, \\
&\dot{x}_{s, m}=\beta\left(x_{s, m}+\epsilon\right)\left(C_{m}-\gamma\right), \\
&\dot{x}_{l, m}=\alpha\left(C_{m}-\delta\right),
\end{aligned}\label{eq:DMM}
\end{equation}
where $x_s\in[0,1]$, $x_l\in[1,10^4M]$, $\alpha=5, \beta=20, \gamma=0.25, \delta=0.05,\epsilon=10^{-3}, \zeta=0.1$. 
The ``gradient-like'' term ($G_{m, i}=\frac{1}{2}q_{m, i}\min(1-q_{m, j}v_j, 1-q_{m, k}v_k)$) and the ``rigidity'' term ($R_{m, i}=\frac{1}{2}(q_{m, i}-v_i)$ if $C_m=\frac{1}{2}(1-q_{m, i}v_i)$, and $R_{m, i}=0$, otherwise) are discussed in Ref.~\onlinecite{seanpaper} along with other details.


\section{Numerical scalability with different integration schemes}
\label{sec:scalability}

\subsection{Numerical details}
To simulate Eq. \eqref{eq:DMM} on modern computers, we need an appropriate numerical integration scheme, which necessitates the discretization of time.  
In Ref.~\onlinecite{seanpaper} we applied the first-order forward Euler method, which, in practice, tends to introduce large numerical errors in the long-time integration. Yet, the massive numerical error in the {\it trajectory} of the ODE solution does not translate into {\it logical} errors in the 3-SAT solution.
The algorithm in Ref.~\onlinecite{seanpaper} was capable of showing {\it power-law} scalability in the typical-case of clause distribution control (CDC) instances~\cite{barthel2002}, though, using an adaptive time step.

To further understand this robustness, we study here the behavior of Eq. \eqref{eq:DMM} under different {\it explicit} numerical integration methods using a {\it constant} time step, $\Delta t$. We will investigate the effect of increasing the time step.
Writing the ODEs as $\dot{\bm{x}}(t)=F(\bm{x}(t))$, with $\bm{x}$ the collection of voltage and memory variables, and 
$F$ the flow vector field~\cite{DMMperspective}, we use the following explicit Runge-Kutta time step~\cite{Sauer}: 

\begin{equation}
    x_{n+1} = x_n + \Delta t \sum_{i=1}^q \omega_i k_i
\end{equation}
where $k_i = F(x_n + \Delta t\sum_{j=1}^{i-1} \lambda_{ij} k_j)$. Specifically, we apply three different integration schemes: {\it forward Euler} method with $q=1$, $\omega_1=1$; {\it trapezoid} method with $q=2$, $\omega = (\frac{1}{2}, \frac{1}{2})$, $\lambda_{21}=1$; {\it 4th order Runge-Kutta } method (RK4) with $q=4$, $\omega=(\frac{1}{6}, \frac{1}{3}, \frac{1}{3}, \frac{1}{6})$, and 
\begin{equation}
    \lambda=\left(\begin{array}{cccc}
       0 & 0 & 0 & 0 \\
       \frac{1}{2} & 0 & 0 & 0 \\
       0 & \frac{1}{2} & 0 & 0 \\
       0 & 0 & 1 & 0
    \end{array}\right).
\end{equation}

{Note that Eqs.~\eqref{eq:DMM} are \emph{stiff}, meaning that the explicit integration methods we consider here should diverge quite fast, within a few integration steps, irrespective of the problem instances we choose~\cite{Sauer}. However, as we will show below, the explicit integration methods actually work very well for these equations, and the integration time step can even be chosen quite large. }

As an illustration, we focus on planted-solution 3-SAT instances at clause-to-variable ratio $\alpha_r=8$~\footnote{Using the method of \cite{barthel2002} with $p_0=0.08$}. These instances require exponential time to solve using the local-search algorithm walk-SAT \cite{barthel2002} and other state-of-the-art solvers \cite{seanpaper}. In the Appendix, we will show similar results for $\alpha_r=6$.

\begin{figure}
\includegraphics[width = 0.48\textwidth]{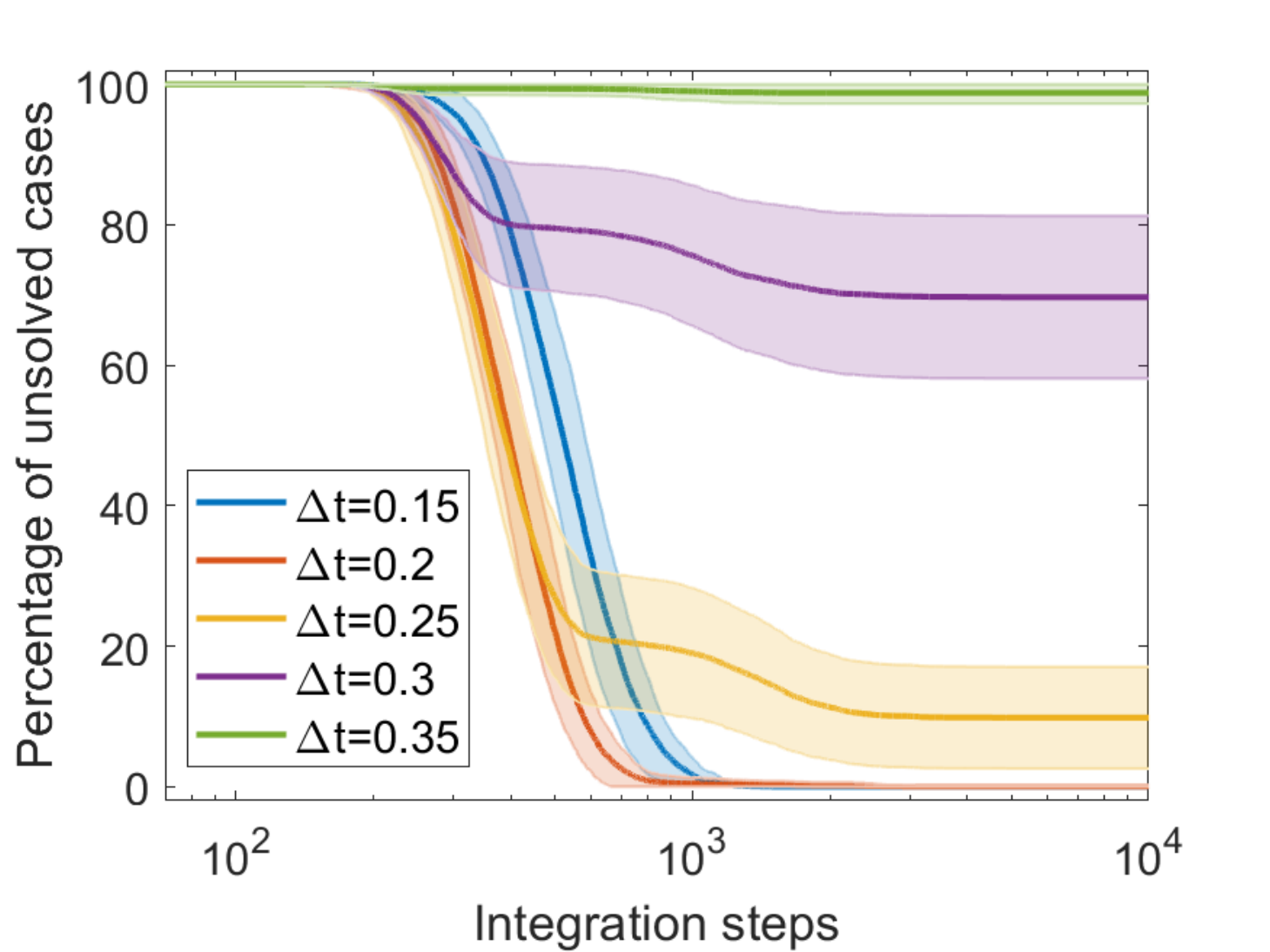}%
\caption{Solution of clause distribution control 3-SAT instances with $10^4$ variables at clause-to-variable ratio $\alpha_r=8$ by numerically integrating Eqs. \eqref{eq:DMM} with the forward Euler method with different time steps, $\Delta t$. We performed $10^4$ solution trials (100 instances and 100 different initial conditions for each instance) for each $\Delta t$. We observe the number of unsolved cases decays as integration steps increase, with the solid line representing the result for all $10^4$ trials, and shaded area representing one standard deviation over the 100 curves calculated separately using the 100 instances, with each curve covering 100 different initial conditions using one specific instance.
	At large number of integration steps, the number of unsolved cases reaches a plateau, whose height only depends on the integration time step $\Delta t$. The fraction of solved instances represents the size of the basin of attraction of Eqs. \eqref{eq:DMM}, which is similar for all instances at a certain $\Delta t$ and shrinks as $\Delta t$ increases. This indicates that rather than being problem-specific, our numerical algorithm is a general incomplete solver for 3-SAT problems.
\label{fig:plateau}}
\end{figure}

\begin{figure}
\includegraphics[width = 0.48\textwidth]{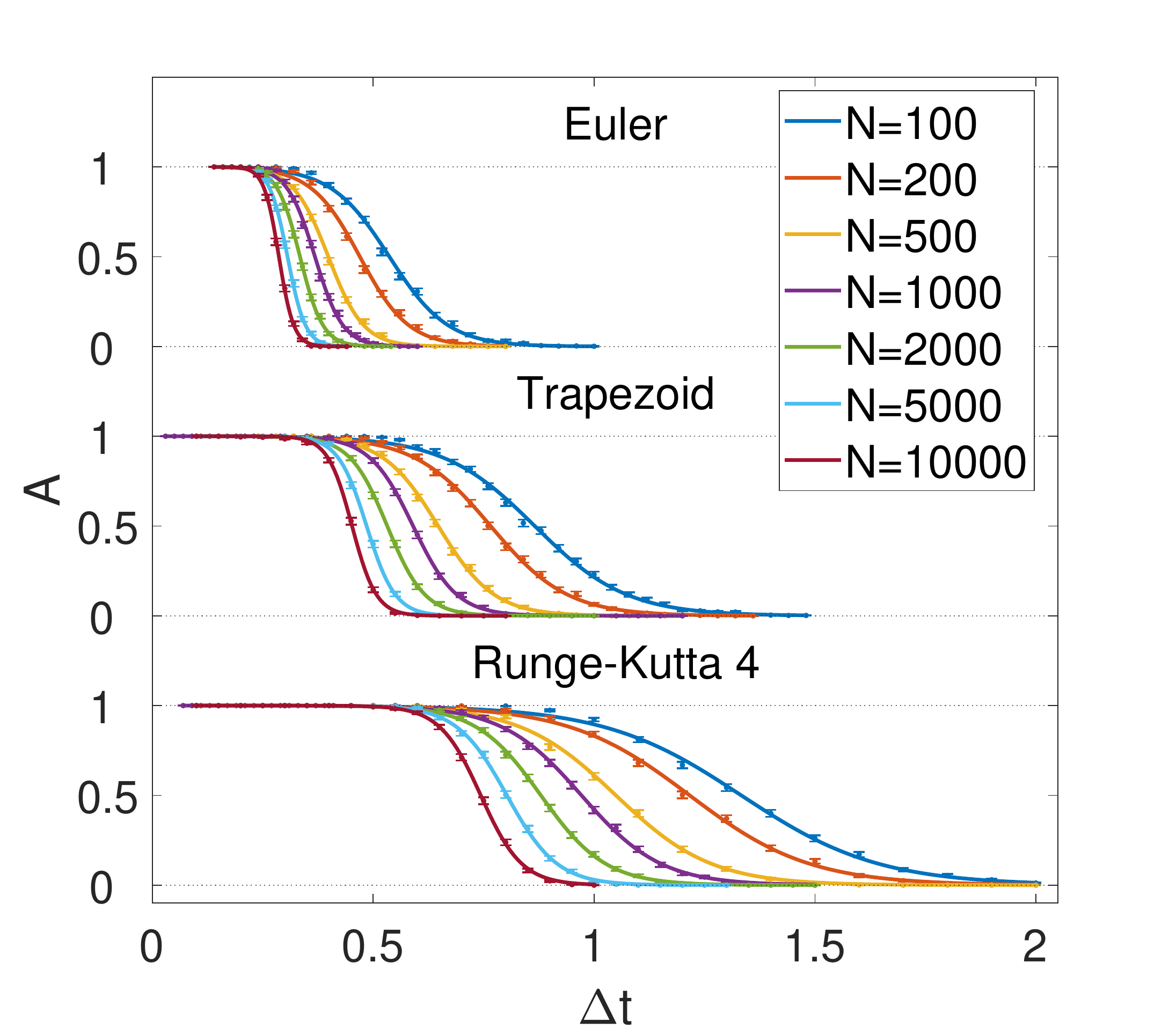}%
\caption{The size of the basin of attraction, $A$, versus $\Delta t$, for different numbers of variables $N$ and different explicit integration schemes, is well 
	fitted by sigmoid-like curves, that become sharper as $N$ increases. This indicates a ``solvable-unsolvable phase transition'' at a critical $\Delta t_c$. Each data point is calculated based on 1000 solution trials (100 instances and 10 initial conditions per instance), and the solid curves are fitted using the function $A=1/[1+\exp(-c(\Delta t-d))]$, with $c$ and $d$ fitting parameters. {The error bars, estimated with the help of data from Fig.~\ref{fig:plateau}, represent one standard deviation. } 
\label{fig:sigmoid_fits}}
\end{figure}

\begin{figure}
\includegraphics[width = 0.48\textwidth]{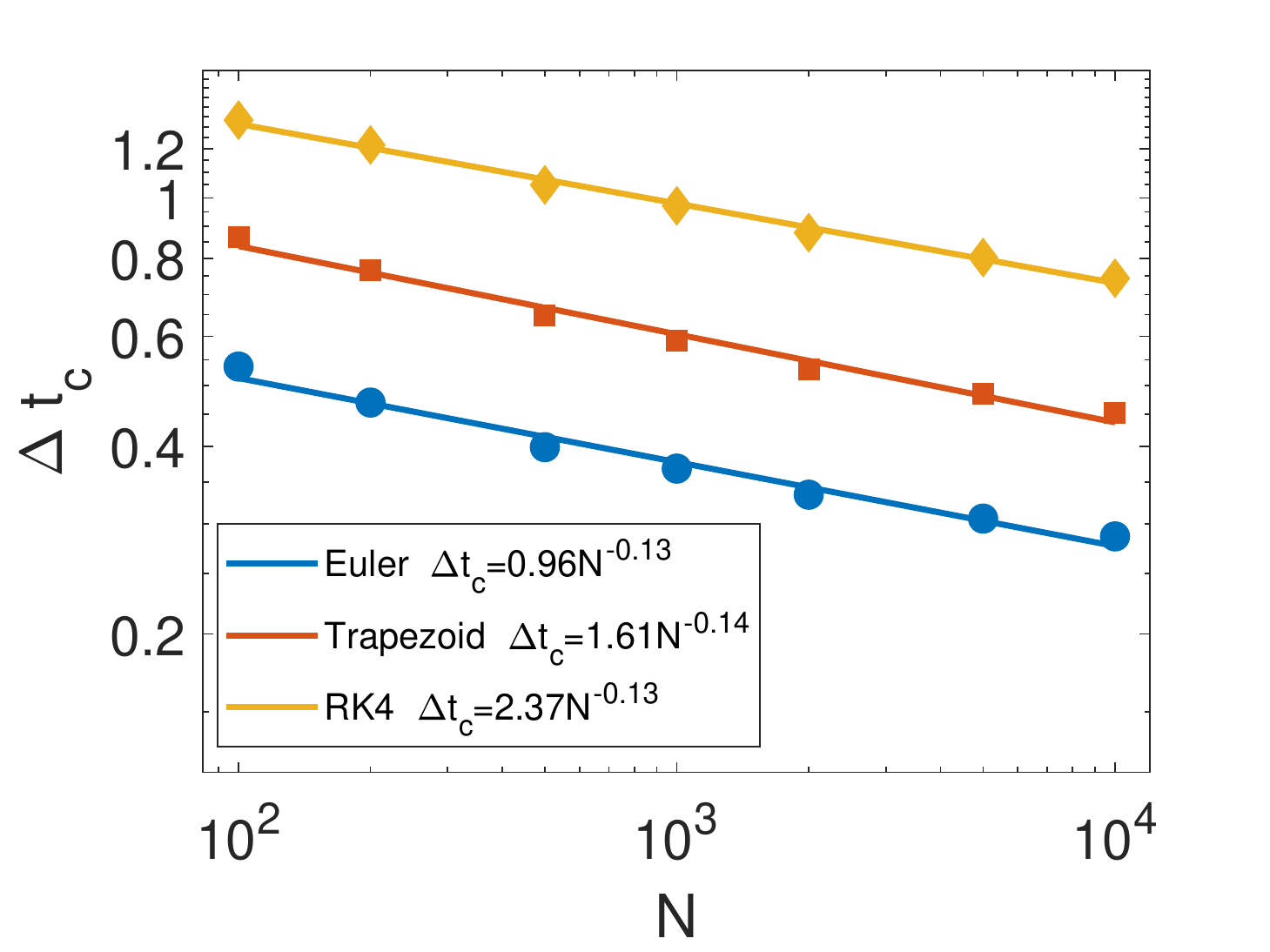}
\caption{In this figure, $\Delta t_c$ is extracted from Fig. \ref{fig:sigmoid_fits} where $A(\Delta t_c)=1/2$. $\Delta t_c$ shows a {\it power-law} scaling  
	with number of variables $N$, and, as expected, integration methods with higher orders have a larger $\Delta t_c$. 
\label{fig:dtc}}
\end{figure}

\begin{figure}
\includegraphics[width = 0.48\textwidth]{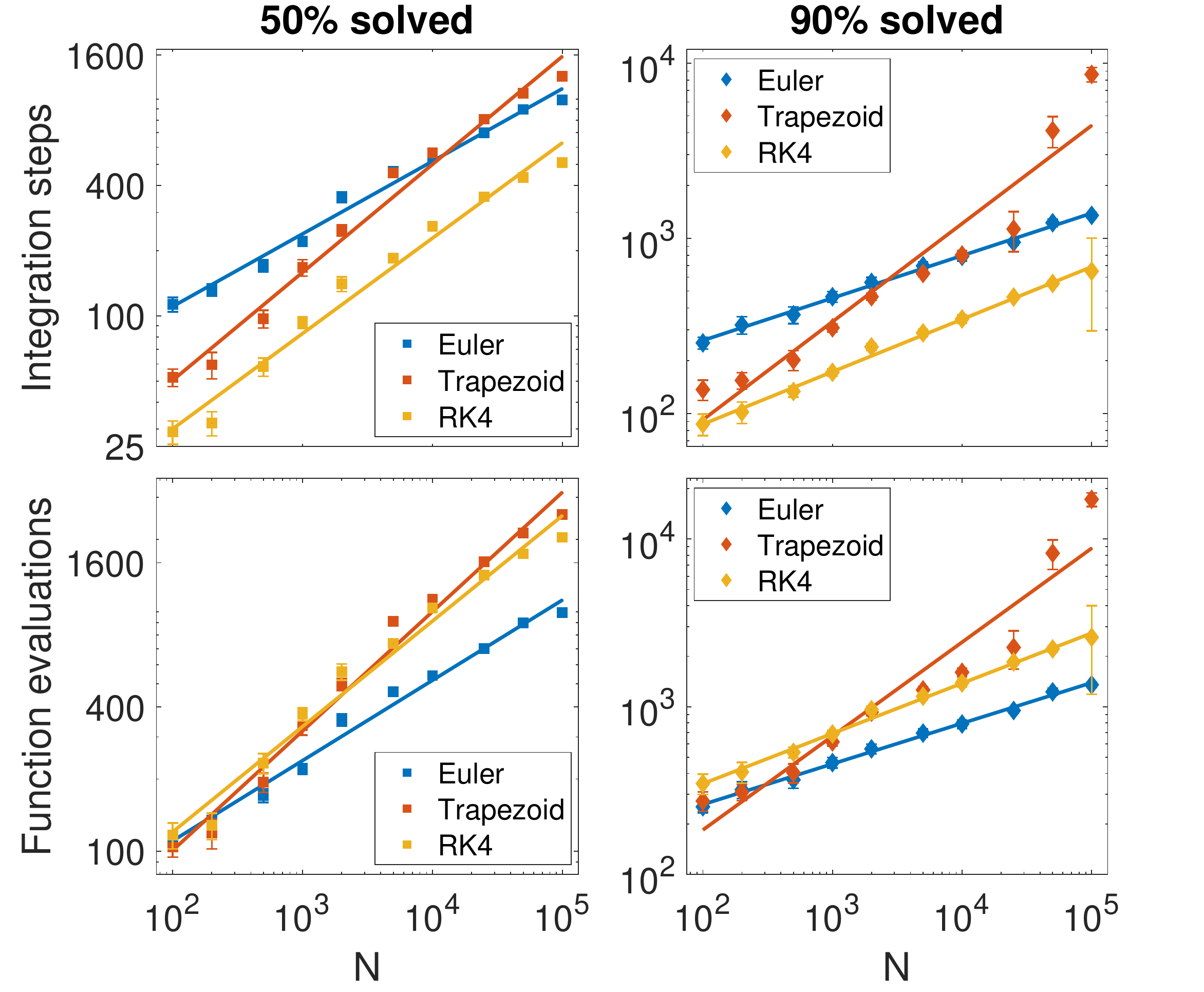}%
\caption{
	The scalability curves for the three different explicit integration schemes considered in this work, in which the integration time step $\Delta t$ is calculated using the power-law fit of $\Delta t$ where $A(\Delta t)=0.95$, times an additional ``safety factor'' of 0.6 (see text). All three methods show a power-law scaling, and the values of the scaling exponents are 0.34 (Euler 50\%), 0.50 (Trapezoid 50\%), 0.44 (RK4 50\%), 0.24 (Euler 90\%), 0.56 (Trapezoid 90\%), 0.30 (RK4 90\%). The scaling is the same, for each integration scheme, in terms of either number of steps or function evaluations; only the pre-factor changes. {The error bars are estimated with bootstrapping and represent one standard deviation.} For $N\in[10^2, 10^5]$, Runge-Kutta 4th order (RK4) requires the least number of integration steps. However, in terms of function evaluations, the simplest forward Euler method is the most efficient one. 
\label{fig:scalability}}
\end{figure}

We start with 100 CDC 3-SAT instances with the number of variables $N=10^4$, and solve them by numerically integrating Eqs. \eqref{eq:DMM} using the forward Euler method with different values of $\Delta t$.  
For each 3-SAT instance, we made 100 individual attempts with different random initial conditions, so that the total number of solution attempts is $10^4$. 
In Fig. \ref{fig:plateau}, we see the number of unsolved instances decreases rapidly until reaching a plateau. When the simulations are performed again with a smaller time step, the plateau height decreases. Once reached, the plateau persists while increasing the number of integration steps.

\subsection{Basin of attraction}
We argue that the plateaus in Fig. \ref{fig:plateau} are caused by a reduction of the basin of attraction in Eqs.~\eqref{eq:DMM}, created by the numerical integration method, rather than the hardness of the instances themselves. 
To show this, first note that {\it all} solution attempts succeed when $\Delta t=0.15$ 
 (basin of attraction is the entire phase space), while almost all attempts fail when $\Delta t=0.35$ (basin of attraction shrinks to zero size). When $\Delta t$ falls in between, for each 3-SAT instance, the number of unsolved attempts is centered around the shaded area  in Fig.~\ref{fig:plateau}. That is, at a certain $\Delta t$, the size of the basin of attraction for different instances is approximately the same. 

We use $A$ to denote the ratio between the volume of the basin of attraction and the volume of the entire phase space. In Fig. \ref{fig:sigmoid_fits}, we estimate this quantity by calculating the fraction of solved instances for each number of variables $N$ and time step $\Delta t$ for 100 different 3-SAT instances per size $N$, and 10 initial conditions per instance. At smaller $\Delta t$, all instances with all different initial conditions are solved, i.e., the basin of attraction is the entire phase space, $A\to 1$. This is in agreement with theoretical predictions for continuous-time dynamics \cite{seanpaper}. On the other hand, when $\Delta t$ is too large, the structure of the phase space gets modified by the numerical error introduced during discretization, and $A\to 0$. This is a well known effect that has been studied also analytically in the literature~\cite{stuart1994numerical}.

Plotting $A$ versus $\Delta t$ for different number of variables $N$ and different integration schemes, we get a series of sigmoid-like curves (Fig. \ref{fig:sigmoid_fits}), that become sharper as $N$ increases. This suggests the existence of a ``phase transition'': when going beyond a critical $\Delta t_c$, there is a drastic change in $A$, and the system undergoes a {\it solvable-unsolvable transition}. 

\subsection{Integration step scalability}
The above results allow us to determine how the integration step, $\Delta t$, for the three integration schemes we have chosen, scales as the size of the problem increases. In Fig. \ref{fig:dtc}, we define $\Delta t_c$ such that $A(\Delta t_c)=1/2$, and determine the relation between $\Delta t_c$ and $N$. 

We find $\Delta t_c$ scales as a {\it power law} with the number of variables $N$. This is a major result because it shows that we do not need $\Delta t$ to decrease exponentially to have the same success rate. In Ref.~\onlinecite{seanpaper}, it was demonstrated, using topological field theory~\cite{topo,DMtopo}, that in the ideal (noise-free) {\it continuous-time} dynamics, the physical time required for our dynamical system to reach the solution scales as $O(N^\gamma)$, with $\gamma\leq 1$ . 

Our numerical results show that, when simulating the dynamics with numerical integration schemes on classical computers, $\Delta t_c\sim N^{-\delta}$. Coupled with the previous analytical results~\cite{seanpaper}, this means that the number of integration steps scales as $O(N^{\gamma+\delta})$. In other words, discretization only adds a $O(N^\delta)$ overhead to the complexity of our algorithm, indicating that DMMs can be efficiently simulated on classical computers~\footnote{Of 
course, this is {\it not} an analytical proof of the efficiency of the {\it numerical} simulations, just an empirical result.}.

In Fig. \ref{fig:dtc}, note that $\Delta t_c > 10^{-1}$ for all three integration methods.
This is quite unexpected for a stiff system simulated with explicit integration methods, because a large time step introduces large local truncation errors at each step of the integration.
This error accumulates and should destroy the {\it trajectory} of the ODEs we are trying to simulate. However, we can still solve {\it all} 
tested instances with Eqs. \eqref{eq:DMM}
even at such a large $\Delta t$. In Sec. \ref{sec:DP}, we will provide an explanation of why this is possible. 

Here, we note that choosing an appropriate $\Delta t$ could speed up the algorithm significantly, as smaller $\Delta t$ leads to excessive integration steps, while larger $\Delta t$ may cause the solver to fail. 
To find an appropriate time step for our scalability tests, we choose $\Delta t$ from Fig.~\ref{fig:sigmoid_fits} such that $95\%$
of the trials have found a solution, and multiply that $\Delta t$ by a ``safety factor'' of $0.6$ to make sure we hit the region of $A(\Delta t)\approx 1$.

By employing the resulting value of $\Delta t$, we plot in Fig. \ref{fig:scalability} the typical-case and 90th percentile scalability curves up to $N=10^5$. For each variable size, $N$, we perform 1000 3-SAT solution trials (100 instances and 10 initial conditions each),
 and report the number of integration steps when $50\%$ and $90\%$ of the trials 
  are solved. Additionally, we report the number of {\it function evaluations} (namely, the number of times the right-hand side of Eqs.~\eqref{eq:DMM} is evaluated), which differs from  the number of integration steps by only a constant. 
  
  Both quantities show a {\it power-law} scalability for all three integration schemes. Since $\Delta t_c$ is larger for higher-ordered ODE solvers, the latter ones typically require fewer steps to solve the problems. However, when taking the total number of function evaluations into account, the simplest forward Euler method becomes the most efficient integration scheme for our ODEs.

\section{Directed percolation and noise in DMMs}
\label{sec:DP}
As anticipated, our results are unexpected in view of the large truncation errors introduced during the simulations of the ODEs 
of the DMMs. However, one might have predicted the numerical robustness upon considering the type of dynamics the DMM executes in phase space. 

\subsection{Instantonic dynamics}       
{Instantons are topologically non-trivial solutions of the equations of motion connecting two different critical points (classical vacua in field-theory language)~\cite{birmingham1991topological}.
In dissipative systems, as the ones considered in this work, instantons connect two critical points (such as saddle points, local minima or maxima), that differ in index (number of unstable directions). In fact, they always connect a critical point with higher index to another with lower index~\cite{Solitons,Coleman}. 
}

It was shown in Refs.~\onlinecite{topo,DMtopo} {that, by stochastically quantizing the DMM's equation of motion, one obtains a supersymmetric topological field theory, from which it can be deduced that} 
the {\it only} ``low-energy'', ``long-wavelength'' dynamics of DMMs is a collection of elementary {\it instantons} (a composite instanton). 
Critical points are also of topological character. For instance, their number cannot change unless we change the topology of phase space~\cite{Fomenko}. 

In addition, given two distinct (in terms of indexes) critical points, there are several (a family of) trajectories (instantons) that may connect them, since the 
unstable manifold of the ``initial'' critical point may intersect the stable manifold of the ``final'' critical point at several points in the phase space. 
In the ideal (noise-free) continuous time dynamics, if the only equilibria (fixed points of the dynamics) are the solutions of the given problem (as shown, e.g., in~\cite{DMM2,seanpaper}), then the state trajectory is {\it driven} towards the equilibria by the voltages that set the input of the problem the 
DMM attempts to solve.

\subsection{Physical noise}
In the presence of (physical) noise instead, {\it anti-instantons} are generated in the system. These are (time-reversed) trajectories that connect critical points 
of {\it increasing} index. However, anti-instantons are {\it gapped}, in the sense that their amplitude is exponentially suppressed with respect to the corresponding 
amplitude of the instantons~\cite{Chaos}. This means that the ``low-energy'', ``long wave-length'' dynamics of DMMs is still a succession of instantons, even in the presence of (moderate) noise.

Nevertheless, suppose an instanton connects two critical points, and immediately after 
an anti-instanton occurs for the {\it same} critical points (even if the two trajectories are different). For all practical purposes, the ``final'' critical point of the original instanton has never been reached, and that critical point can be called {\it absorbing}, in the sense that the trajectory would ``get stuck'' on that one, or wanders in other regions of the phase space. In other words, 
in the presence of noise, there is a finite probability for some state trajectory to explore a much larger region of the phase space. The system then needs to explore some other (anti-)instantonic trajectory to reach the equilibria, solutions of the given problem. Nevertheless, due to the fact that anti-instantons are gapped, and the topological character of both critical points and instantons connecting them, if the physical noise level is not high enough 
to change the topology of the phase space, the dynamical system would still reach the equilibria.~\footnote{In fact, moderate noise may even help {\it accelerate} the 
time to solution, by reducing the time spent on the critical points's stable directions.} 

{On the other hand, if the noise level is too high, our system may experience a condensation of instantons and anti-instantons, which can break supersymmetry dynamically~\cite{Entropy}. In turn, this supersymmetry breakdown, can produce a noise-induced chaotic phase even if the flow vector field remains integrable. Although we do not have an analytical proof of this yet, we suspect that this may be related to the unsolvable phase we observe.}

\subsection{Directed percolation}
If we visualize the state trajectory as the one traced by a liquid in a corrugated landscape, the above suggests an intriguing analogy with the phenomenon of {\it 
directed percolation} (DP)~\cite{henkel2008non}, with the critical points acting as `pores', and instantons as `channels' connecting `neighboring' 
(in the sense of index difference) critical 
points. 

DP is a well-studied model of a non-equilibrium (continuous) phase transition from a fluctuating permeable phase to an absorbing phase \cite{henkel2008non}. It can be intuitively understood as a liquid passing through a porous substance under the influence of a field (e.g., gravity). 
The field restricts the direction of the liquid's movement, hence the term ``directed''. In this model, neighboring pores are connected by channels with probability $p$, and disconnected otherwise. When increasing $p$, the system goes through a phase transition from the absorbing phase into a permeable phase at a critical threshold $p_c$. 

\subsection{Numerical noise}
In numerical ODE solvers, discrete time also introduces some type of noise (truncation errors), and the considerations we have 
made above on the presence of anti-instantons still hold. However, numerical noise could be substantially more 
damaging than physical noise. The reason is twofold. 

First, as we have already anticipated, numerical noise {\it accumulates} during the integration of the ODEs. In some sense, it is then always {\it non-local}. Physical noise, instead, is typically local (in space and time), hence it is a {\it local} perturbation of the dynamics~\cite{vanKampen}. As such, if it is not too large, it cannot change the phase space topology. 

Second, unlike physical noise, integration schemes may change the topology of phase space {\it explicitly}. This is because, when one transforms the 
continuous-time ODEs~\eqref{eq:DMM} into their discrete version (a {\it map}), this transformation can introduce {\it additional} critical points in the phase space of the map, which were not present in the original phase space of the continuous dynamics. These extra (undesirable) critical points are sometimes called {\it ghosts}~\cite{cartwright1992dynamics}. For instance, while the forward Euler method can {\it never} introduce such ghost critical points, irrespective of the size of $\Delta t$ (because both the continuous dynamics and the associated map have the same flow field $F$), 
both the trapezoid and Runge-Kutta 4th order may do so if $\Delta t$ is large enough. These critical points would then further degrade the dynamics 
of the system.

With these preliminaries in mind, let us now try to quantify the analogy between the DMM dynamics in the presence of numerical noise and directed percolation.

\subsection{Paths to solution}
\label{sec:paths}
First of all, the integration step, $\Delta t$, must be inversely related to the percolation probability $p$: when $\Delta t$ tends to zero, the discrete dynamics approach  
the ideal (noise-free) dynamics for which $p\rightarrow 1$. In the opposite case, when $\Delta t$ increases, $p$ must decrease. 

In directed percolation, the order parameter is the density of active sites~\cite{henkel2008non}. This would correspond to the density of ``achievable'' critical points in DMMs. However, due to the vast dimensionality of their phase space (even for relatively small problem instances), this quantity is not directly accessible in DMMs.

Instead, what we can easily evaluate is the ratio of successful solution attempts: starting from a random point in the phase space (initial condition of the ODEs~\eqref{eq:DMM}), and letting the system evolve for a sufficiently long time, a path would either successfully reach the solution 
(corresponding to a permeable path in DP) 
or fail to converge 
(corresponding to an absorbing path in DP)
. By repeating this numerical experiment for a large number of trials, the starting points of the permeable paths essentially fill the basin of attraction of our dynamical system. The advantage of considering this quantity is that the ratio of permeable paths in bond DP models can be calculated analytically. 

\begin{figure}[!t]
\includegraphics[width = 0.48\textwidth]{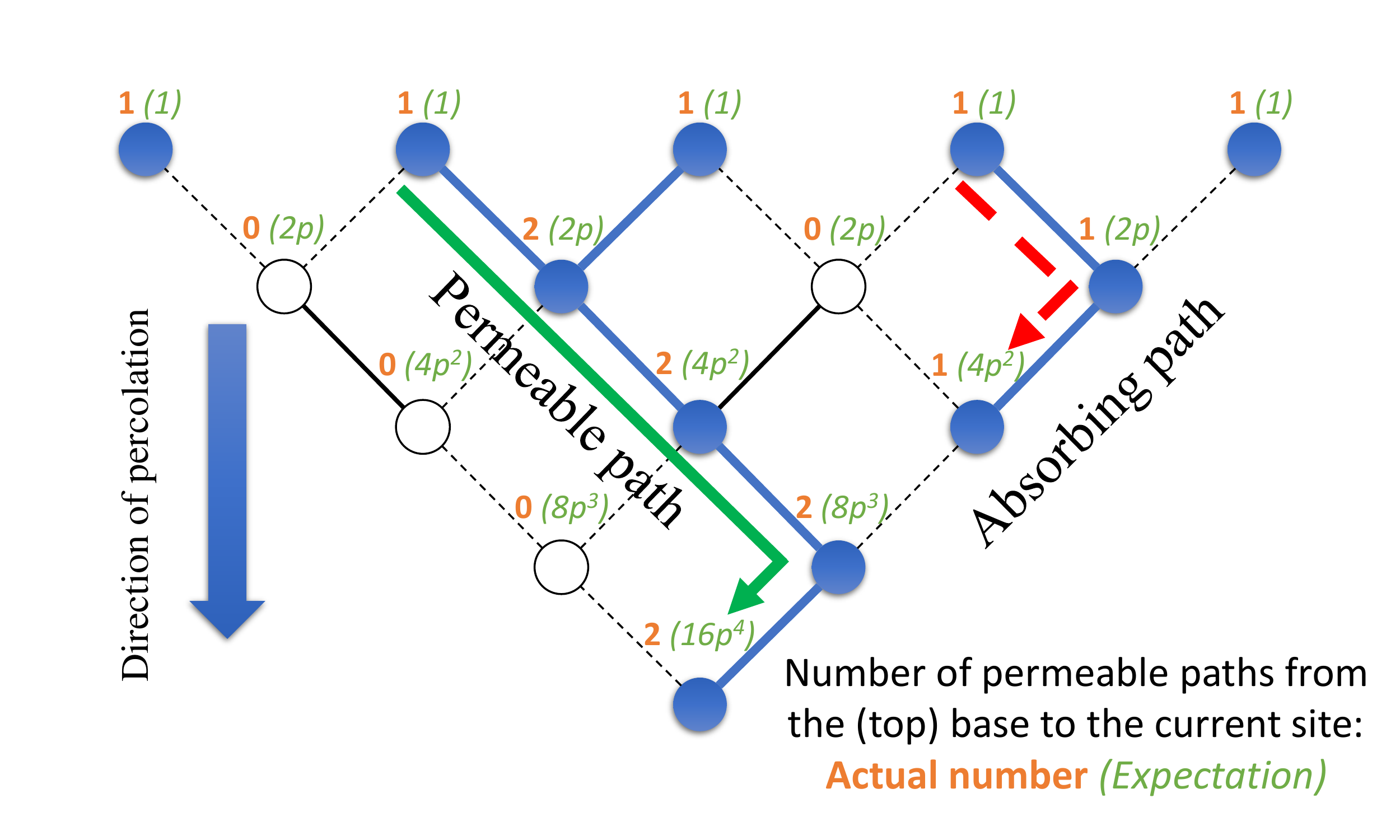}%
\caption{Illustration of permeable and absorbing paths in DP on a cone-shaped 2d lattice. Connected bonds are represented with solid lines, and disconnected ones are represented with dotted lines. A permeable path connects the base to the apex, and an absorbing path ends in the middle of the lattice with all bonds beneath it disconnected. The number of permeable paths $n_\mathrm{p}$ is calculated iteratively: $n_\mathrm{p}$ at a given site equals the sum of $n_\mathrm{p}$ at its connected predecessor sites, and the expectation $\langle n_\mathrm{p}\rangle$ at a given site equals the sum of $n_\mathrm{p}$ at all its predecessor sites times $p$, the percolation probability.
\label{fig:DP_illustration}}
\end{figure}

{Now we set up the correspondence of paths between DP and DMM. In DP, a permeable path is defined to be a path that starts from the top of the lattice and ends at the bottom, and corresponds to a successful solution attempt in DMMs. Similarly, an absorbing path starts from the top and terminates within the lattice in DP, and corresponds to a failed solution attempt in DMMs.}
Consider then a 3-SAT problem with only one solution. A DMM for such a problem can reach the solution from several initial conditions. This translates into a $D$-dimensional cone-shaped lattice for the 
DP model (see Fig. \ref{fig:DP_illustration}). The (top) base of the cone would represent the starting points, and the apex of the cone would represent the solution point. A permeable path connects the base to the apex, and an absorbing path ends in the middle of the lattice with all bonds beneath it disconnected. 

{Note that, in the DP model, it is possible to have different permeable paths starting from the same initial point. However, from the perspective of DMMs, there is no randomness in the dynamics, and each initial condition corresponds to a unique path. Since DP is a simplified theoretical model to describe numerical noise in DMMs, in this model we assume that the trajectory ``randomly chooses'' a direction when reaching a critical point. This is a reasonable assumption since there can be a large number of instantons starting from some critical point in the phase space, and it is almost impossible for us to actually monitor which instantonic path the trajectory follows. }

{With this correspondence in mind, the ratio of permeable paths in DP is analogous to the ratio of successful solution attempts in DMMs. Below, we will calculate this quantity exactly in DP, and use this calculation to model the solvable-unsolvable transition we found in DMMs. }

To begin with, we assume that all starting points are occupied with equal probability, and calculate the expectation of the number of permeable and absorbing paths. This can be done iteratively: the expected number of permeable paths at a given site equals the sum of permeable paths at its predecessor sites, times $p$, the percolation probability. Assuming the number of time steps required to reach the apex is $T$, then the expected number of permeable paths
\begin{equation}
    \langle n_{\mathrm{p}, T}\rangle=(Dp)^T.
\end{equation}
An illustration of the calculation on a cone-shaped 2d lattice is shown in Fig.~\ref{fig:DP_illustration}.

\begin{figure}[!t]
\includegraphics[width = 0.48\textwidth]{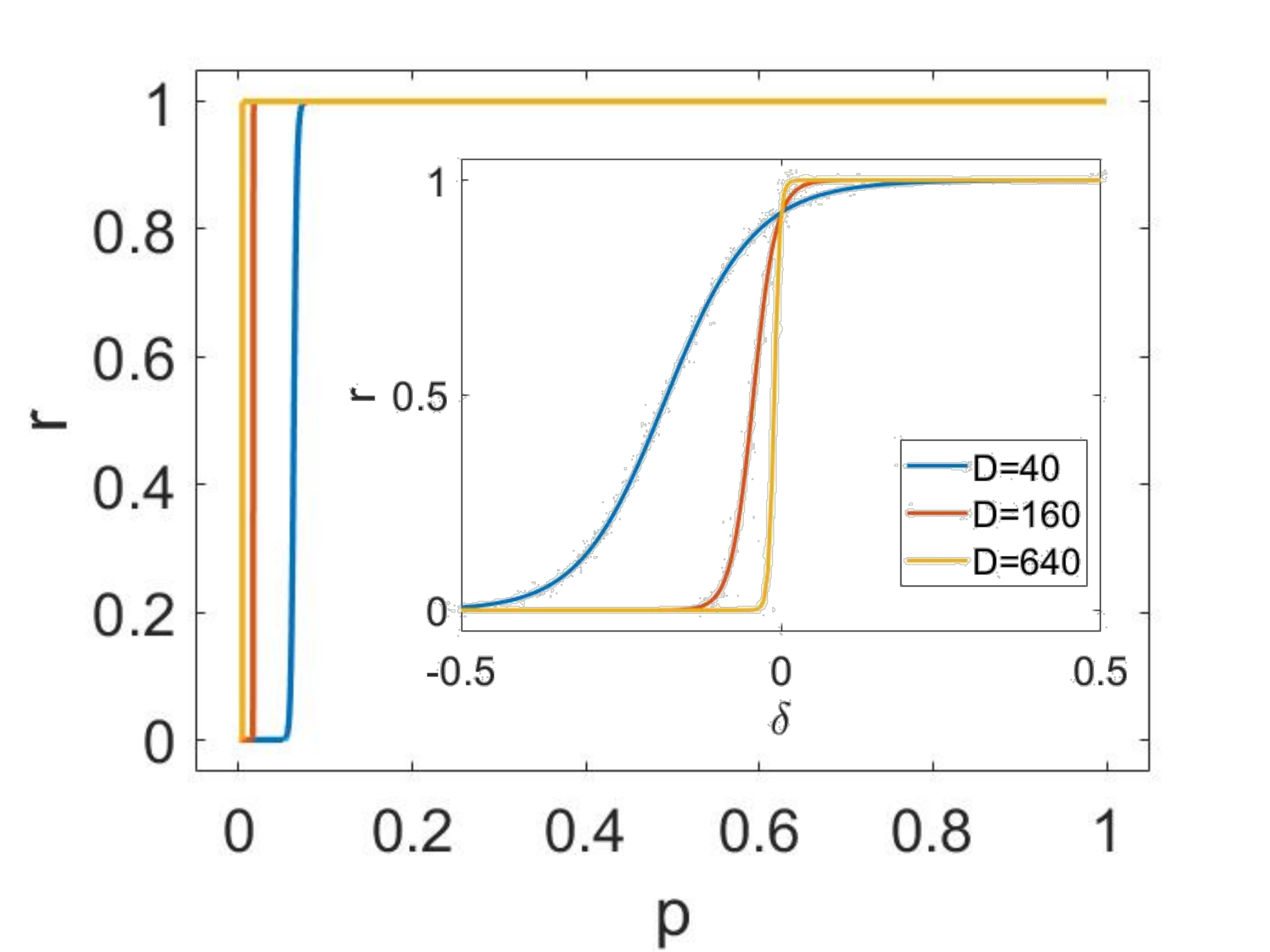}%
\caption{
{The ratio of permeable paths, calculated with Eq.~\eqref{eq:ratio}, plotted for different values of the dimension $D$. The inset shows the same curves near their transition points, with $\delta=Dp-e$.} The curves exhibit a sigmoidal behavior, which is similar to the one in Fig.~\ref{fig:sigmoid_fits}. However, note that, since $\Delta t$ in DMMs is inversely related to the probability $p$ in DP, they are curving in opposite directions.
\label{fig:r_theory}}
\end{figure}

{The number of absorbing paths $\langle n_{a, T}\rangle$ is trickier to compute. Details of the calculation can be found in Appendix~\ref{AppendixB}, here we only give the approximate result (valid for $Dp>1$) near the transition:
\begin{equation}
    \langle n_{\mathrm{a}, T}\rangle=\frac{(Dp)^{T+1}}{2}\left(\frac{1-p}{\ln Dp}\right)^D\erfc(\sqrt{D}\frac{1-\ln Dp}{\sqrt{2\ln Dp}}).
    \label{eq:na}
\end{equation}

The ratio of permeable paths $r$ is simply $\langle n_{\mathrm{p}, T}\rangle/(\langle n_{\mathrm{p}, T}\rangle+\langle n_{\mathrm{a}, T}\rangle)$: 
\begin{equation}
    r=\frac{1}{1+\frac{1}{2} Dp(\frac{1-p}{\ln Dp})^D\erfc(\sqrt{D}\frac{1-\ln Dp}{\sqrt{2\ln Dp}})}.
    \label{eq:ratio}
\end{equation}

We observe that the transition occurs at $p_c\sim \frac{e}{D}$. Near the transition, let us define $p=\frac{e+\delta}{D}$, where $\delta$ is small. Then, $\ln Dp\approx 1+\frac{\delta}{e}$. Further using $\lim_{x\to\infty}(1+\frac{1}{x})^x=e$, to order $O(\delta)$,  Eq.~\eqref{eq:ratio} becomes

\begin{equation}
\begin{aligned}
    r=\frac{1}{1+\frac{1}{2}e^{1-e-\delta-D\delta/e}\erfc\big(-\sqrt{\frac{D}{2}}\frac{\delta}{e}\big)}.
\end{aligned}\label{eq:ratio_approx}
\end{equation}

In the limit of $D\to\infty$, the divergence comes from the $D$ in the exponent. Therefore, the transition happens exactly at $\delta=0$. When $\delta>0$, $r\to 1$; when $\delta<0$, $r\to 0$. 

Note that when $\delta\ll 0$, the $\erfc$ term dominates over the $e^{-D\delta/e}$ term instead. However, in this case, $p$ is too small, and some approximations we made in Appendix~\ref{AppendixB} to derive $\langle n_{\mathrm{a}, T}\rangle$ no longer hold. In this sense, Eqs.~\eqref{eq:ratio} and \eqref{eq:ratio_approx} are only valid near the transition point $p_c=\frac{e}{D}$. 

The behavior of Eq.~\eqref{eq:ratio} is plotted in Fig.~\ref{fig:r_theory} for different dimensions $D$.} Comparing to Fig.~\ref{fig:sigmoid_fits}, we can already see a few similarities: both figures exhibit a sigmoidal behavior. However, as $\Delta t$ in DMMs is inversely related to $p$ in DP, they are curving in opposite directions. 



Recall that the size of the basin of attraction, $A$, in DMMs corresponds to the ratio $r$ in DP. To model their relation, we then use the {\it ansatz} {$\delta\equiv Dp-e=a\left(\frac{1}{N\Delta t}-b\right)$}, with $a$ and $b$ some real numbers, and fit $A$ to $\Delta t$ using Eq. \eqref{eq:ratio}. (Note that this trial function has a meaning only near $\Delta t_c$.)

\begin{figure}
\includegraphics[width = 0.48\textwidth]{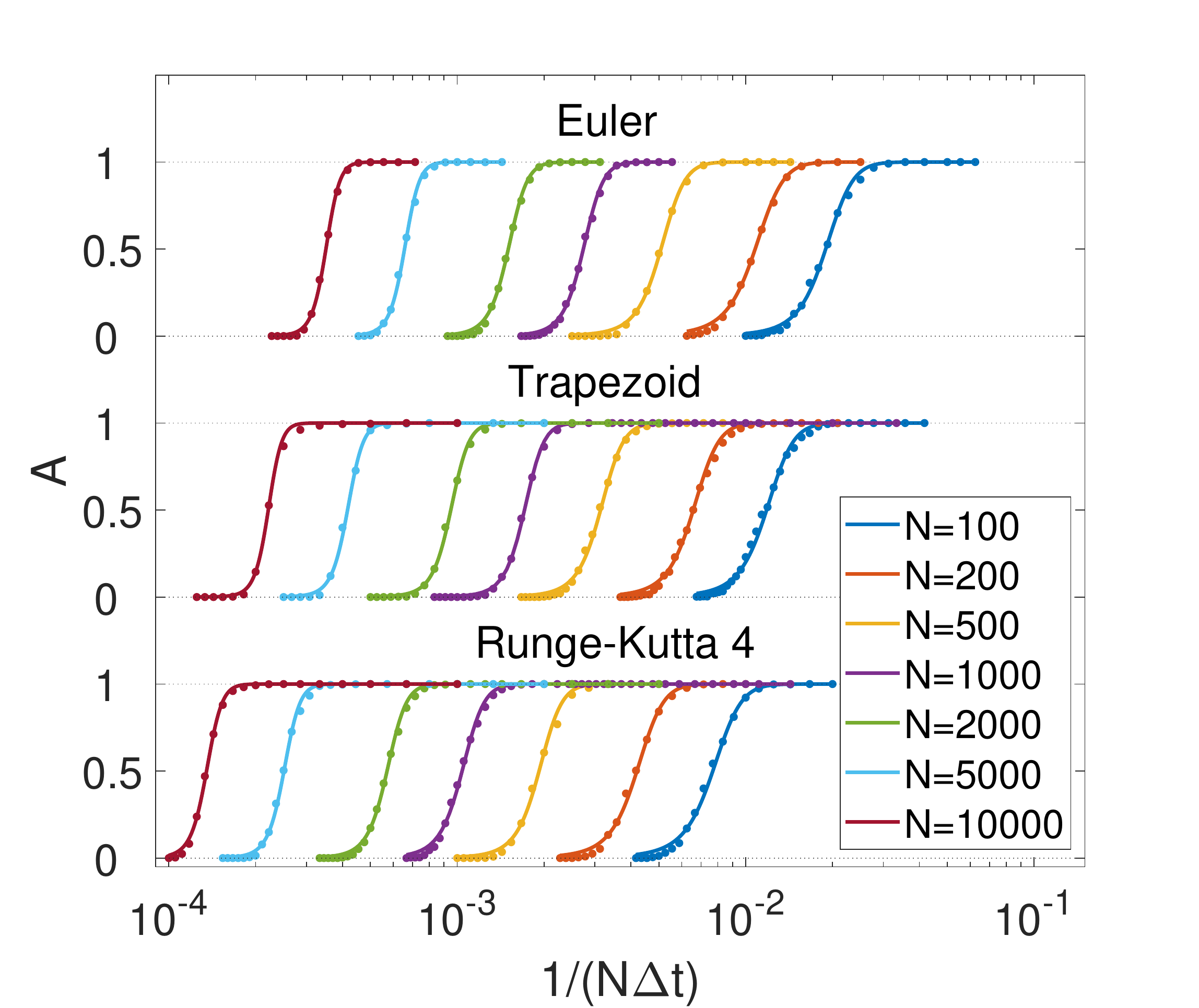}%
\caption{
Relation, close to $\Delta t_c$, between the size of the basin of attraction $A$ and the discretization time step $\Delta t$, plotted for different system sizes $N$, for the three explicit methods used in this work. Each data point is obtained by numerically integrating Eqs. \eqref{eq:DMM} over 1000 3SAT solution attempts (100 instances and 10 initial conditions each), until reaching a plateau. The solid curves are fitted using Eq. \eqref{eq:ratio}, with $A$ corresponding to $r$ and { $\delta\equiv Dp-e=a\left(\frac{1}{N\Delta t}-b\right)$}, with $a$ and $b$ being fitting parameters. 
\label{fig:theoretic_fit}}
\end{figure}

\begin{figure}
\includegraphics[width = 0.48\textwidth]{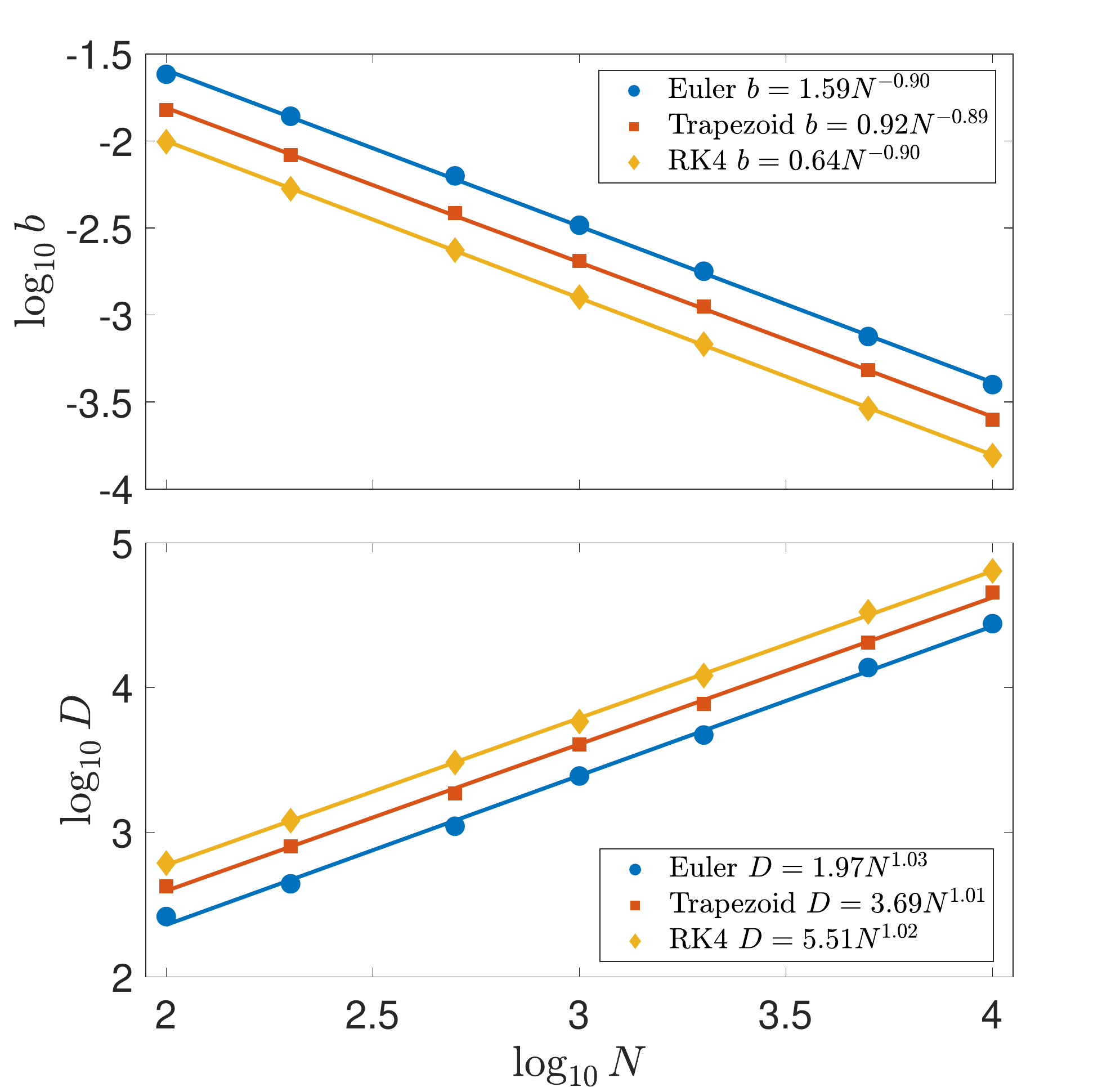}%
\caption{ 
{The fitted parameters $b$ and $D$ for different variable size $N$ for the trial function $\delta\equiv Dp-e=a\left(\frac{1}{N\Delta t}-b\right)$, which connects the time step $\Delta t$ to the percolation probability $p$. ($a$ is fixed to 5.)} $D$ is the dimensionality of the DP lattice, which corresponds to the dimensionality of the composite instanton. 
\label{fig:fit_params}}
\end{figure}


{We find that we can fit the curves reasonably well by fixing $a=5$, and Fig. \ref{fig:theoretic_fit} shows the fitting result of $b$ and $D$}, where the number of variables $N$ ranges between $10^2$ and $10^4$, and each data point is obtained by numerically integrating Eqs. \eqref{eq:DMM} for 1000 3-SAT solution attempts (100 instances and 10 initial conditions each),
until a plateau, as in Fig. \ref{fig:plateau}, has been reached.  Note that the horizontal axis represents $1/(N\Delta t)$.

{
The fitted parameters $b$ and $D$ are shown in Fig.~\ref{fig:fit_params}, where both parameters exhibit a power-law scaling. At the transition threshold $\delta=0$, we have $b=\frac{1}{N\Delta t}$. Therefore, the power-law scaling of $b$ is closely related to the power-law scaling of $\Delta t_c$ in Fig.~\ref{fig:dtc}, and these two scalings show similar trends.
}



The dimensionality $D$ is proportional to the dimensionality of the composite instanton (namely the number of elementary instantons to reach the solution). In Ref.~\onlinecite{DMtopo}, this dimensionality was predicted to scale (sub-)linearly with the number of variables $N$ (in the noise-free case), and it is smaller than the dimensionality of the actual phase space ($(1+2\times \alpha_r)N=17N$ in the present case). This prediction agrees with our numerical results. 

Finally, using the correspondence between DMMs in the presence of noise and DP, we can qualitatively explain why the DMMs numerically integrated still 
provide solution to the problem they are designed to solve even with such large numerical errors introduced during integration. In DMMs, the dimensionality of the composite instanton, $D$, is usually very large, and DP tells us that the percolation threshold {$p_c=e/D\sim e/N$}.

Therefore, even if most of the paths in the phase space are destroyed by noise, we can still achieve the solution {as long as the probability of an instantonic path is larger than $e/D\sim e/N$.} This argument, in tandem with the fact that critical points and instantons have a topological character~\cite{Solitons,Coleman}, ensures the robustness of DMMs in numerical simulations.

\section{Conclusions}
\label{sec:conclusion}

In conclusion, we have shown that the dynamics of digital memcomputing machines (DMMs) under discrete numerical solvers can be described as a directed percolation of state trajectory in the phase space. The inverse time step, $1/\Delta t$, plays the role of percolation probability, and the system undergoes a ``solvable-unsolvable phase transition'' at a critical $\Delta t_c$, which scales as a {\it power law} with problem size. In other words, for the problem instances considered, we have numerically found that the integration time step does {\it not} need to decrease exponentially with the size of the problem, in 
order to control the numerical errors. 

This result is quite remarkable  
considering the fact that we have only employed {\it explicit} methods of integration (forward Euler, trapezoid, and Runge-Kutta 4th order) for {\it stiff} ODEs. (In fact, the forward Euler method, although having the largest numerical error, solves the instances in the least amount of function evaluations.) It can be ultimately traced to the type of dynamics the DMMs perform during the solution search, which is a composite instanton in phase space. Since instantons, and the critical points they connect, are of topological character, perturbations to the actual trajectory in phase space do not have the same detrimental effect as the changes of topology of the phase space. 

Since numerical noise is typically far worse 
than physical noise, these results further reinforce the notion that these machines, if built in hardware, would be topologically protected against 
moderate physical noise 
and perturbations. 

However, let us note that we did not prove that the dynamics of DMMs with numerical (or physical) noise belong to the DP universality class. In fact, according to the DP-conjecture \cite{janssen1981nonequilibrium, grassberger1982phase}, a given model belongs to such a universality class if {\it (i)} the model displays a continuous phase transition from a fluctuating active phase into a unique absorbing state; {\it (ii)} the transition is characterized by a non-negative one-component order parameter; {\it (iii)} the dynamic rules are short-ranged; {\it (iv)} the system has no special attributes such as unconventional symmetries, conservation laws, or quenched randomness.

It is easy to verify that DMMs under numerical simulations satisfy properties {\it (iii)} and {\it (iv)}. However, verifying property {{\it (i)} and} {\it (ii)} is not trivial, {as the relevant order parameters are not directly accessible due to }
the vast phase space of DMMs. Still, the similarities we have outlined in this paper, between DMMs in the presence of noise and DP, help us better understand how these dynamical systems with memory work, and why their simulations are robust against the unavoidable numerical errors. 

\section*{Data availability}
The 3-SAT instances and raw data used to generate all figures in this paper are available upon request from the authors.

\section*{Acknowledgments}
We thank Sean Bearden for providing the MATLAB code to solve the 3-SAT instances with DMMs, as well as insightful discussions. Work supported by NSF under grant No. 2034558. All memcomputing simulations reported in this paper have been done on a single core of an 
AMD EPYC server. 

\clearpage
\setcounter{equation}{0} \makeatletter
\renewcommand \theequation{A\@arabic\c@equation}
\renewcommand \thetable{A\@arabic\c@table}

\setcounter{secnumdepth}{3}

\appendix

\section{Results for $\alpha_r=6$}\label{AppendixA} 

Here, we show that the results presented in the main text hold also for other clause-to-variable ratios. As examples, we choose $\alpha_r=6$. (Similar scalability results have been already reported in Ref.~\onlinecite{seanpaper} for $\alpha_r=4.3$.) In particular, we find similar scaling behavior for $\Delta t_c$ for all clause-to-variable ratios, with of course, different power laws. 

\begin{figure}[htbp]
	\includegraphics[width = 0.48\textwidth]{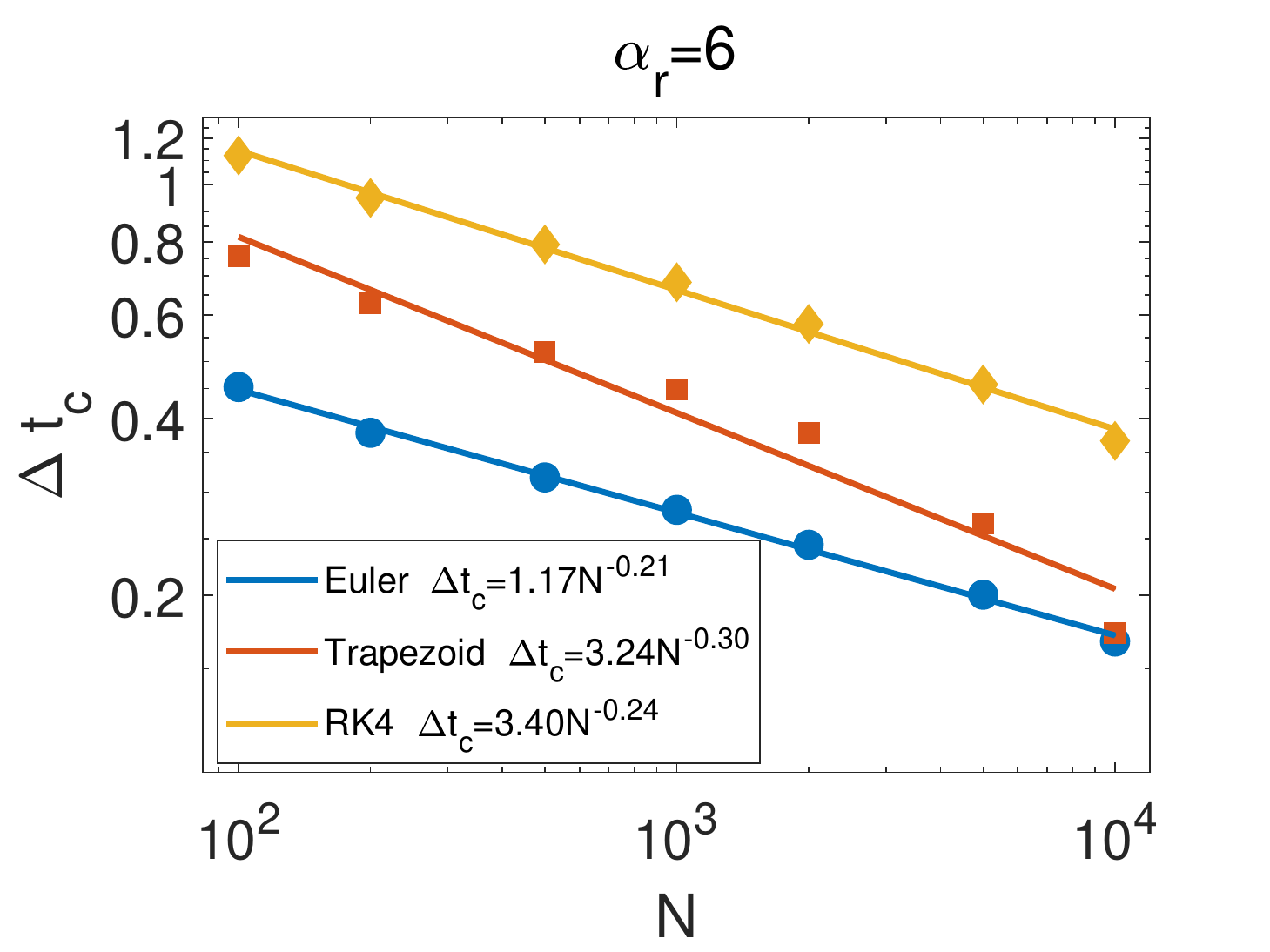}
	\caption{Critical $\Delta t_c$ defined as $A(\Delta t_c)=1/2$ (see main text) for a clause-to-variable ration of $\alpha_r =6$. $\Delta t_c$ shows a {\it power-law} scaling  
		with $N$, and, as expected, integration methods with higher orders have a larger $\Delta t_c$. 
		\label{fig:dtc_ratio_6}}
\end{figure}

Figure~\ref{fig:dtc_ratio_6} have been obtained as discussed in the main text, and show $\Delta t_c$ vs. number of variables 
for $\alpha_r=6$.  

\begin{figure}[htbp]
	\includegraphics[width = 0.48\textwidth]{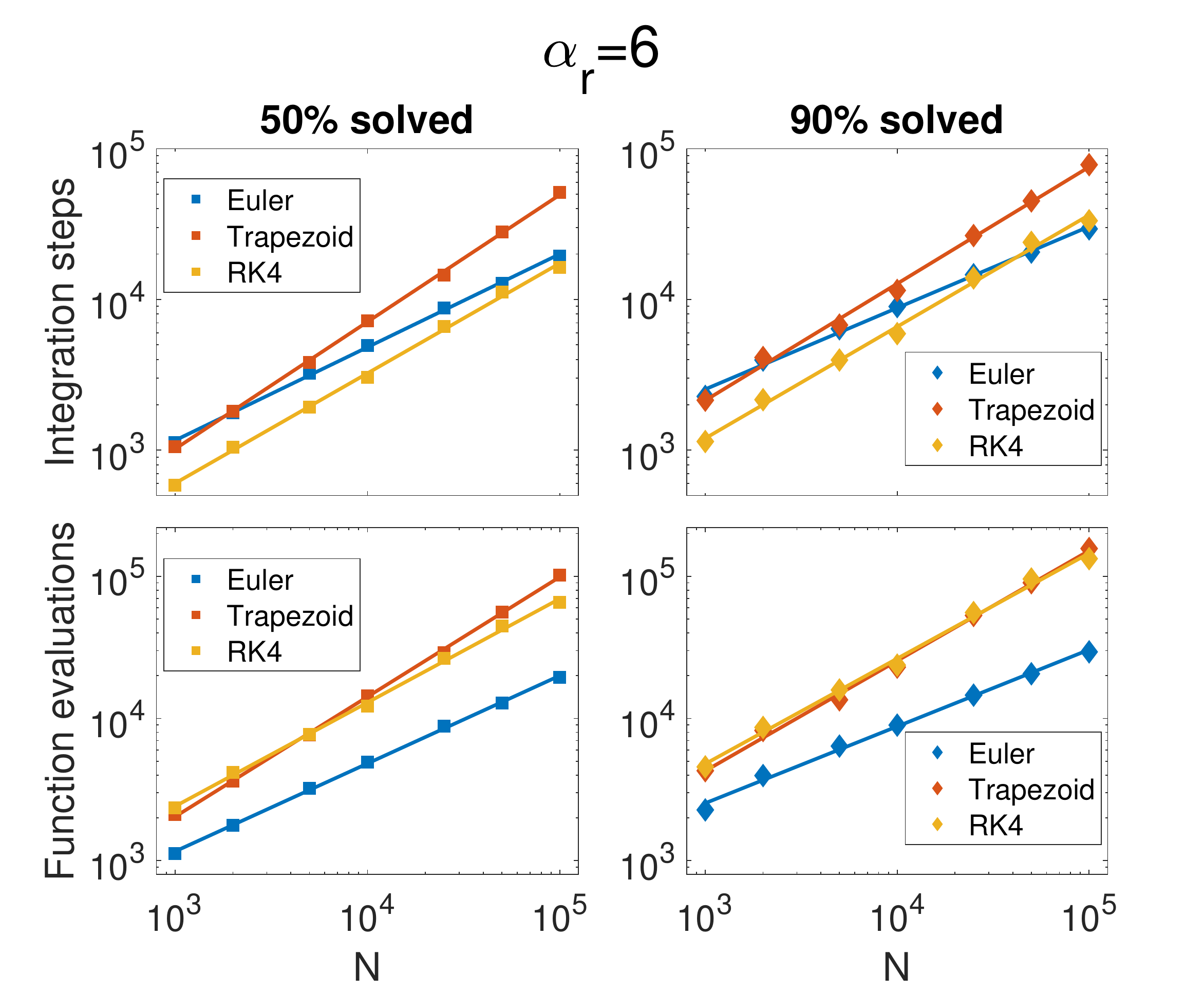}%
	\caption{
		The scalability curves at $\alpha_r=6$ for the three different explicit integration schemes considered in this work, in which the constant integration time step $\Delta t$ is calculated using the power-law fit of $\Delta t$ where $A(\Delta t)=0.95$, times an additional ``safety factor'' of 0.6 (see main text). All three methods show a power-law scaling. The values of the scaling exponents are 0.62 (Euler 50\%), 0.84 (Trapezoid 50\%), 0.73 (RK4 50\%), 0.54 (Euler 90\%), 0.77 (Trapezoid 90\%), 0.74 (RK4 90\%). The scaling is the same, for each integration scheme, in terms of either number of steps or function evaluations; only the pre-factor changes. 
		For $N\in[10^3, 10^5]$, Runge-Kutta 4th order (RK4) requires the least number of integration steps. However, in terms of function evaluations, the simplest forward Euler method is the most efficient one. 
		\label{fig:scalability_ratio_6}}
\end{figure}

In Fig.~\ref{fig:scalability_ratio_6}, instead we show the scalability curves for $\alpha_r=6$, considering all three explicit integration methods. 
As in the main text, we find that the forward Euler method, although having the largest numerical error, solves the instances in the least amount of function evaluations for $\alpha_r=6$. 
Every data point in the curves is obtained with 100 3-SAT instances, with 10 solution trials for each instance. 

\setcounter{equation}{0} \makeatletter
\renewcommand \theequation{B\@arabic\c@equation}
\renewcommand \thetable{B\@arabic\c@table}

\section{Calculating the number of absorbing paths in directed percolation}
\label{AppendixB}
Here, we give a detailed calculation of the number of absorbing paths we outlined in Sec.~\ref{sec:paths}. We use $D$ to denote the dimension of the lattice where percolation takes place (corresponding to the dimension of the DMM phase space), $T$ to denote the number of steps to reach the bottom of the lattice (corresponding to the number of instantonic steps to reach the solution in DMMs), and $p$ to denote the percolation probability. Throughout the calculation, we use the approximation that $D\to\infty, T\to\infty$, as both $D$ and $T$ are very large in DMMs. 

As we illustrated in Sec.~\ref{sec:paths}, the expected number of permeable paths at time step $i$ (i.e., the $i$-th level in the lattice in Fig.~\ref{fig:DP_illustration}, where $i$ starts from 0), is
\begin{equation}
    \langle n_{\mathrm{p}, i}\rangle=(Dp)^i
\end{equation}

The number of absorbing paths at each site is the number of permeable paths at that site, times $(1-p)^D$, the probability that all bonds beneath it are disconnected. Note that this probability has a different expression at the boundary of the lattice, but as $D$ and $T$ are large, we ignore the boundary effect here. Meanwhile, the number of sites $m_i$ at each time step $i$, to the leading order, equals to the volume of a $(D-1)$-dimensional hyperpyramid, 
\begin{equation}
    m_i = \frac{(T+1-i)^{D-1}}{(D-1)!}.
\end{equation}

Then, the total number of absorbing paths is
\begin{equation}
\begin{aligned}
    &\langle n_{\mathrm{a}, T}\rangle = (1-p)^D\sum_{i=0}^{T-1} \frac{(T+1-i)^{D-1}}{(D-1)!}(Dp)^i\\
    &= \frac{(1-p)^D (Dp)^{T+1}}{(D-1)!}\sum_{i=0}^{T-1}(T+1-i)^{D-1}\frac{1}{(Dp)^{T+1-i}}\\
    &= \frac{(1-p)^D (Dp)^{T+1}}{(D-1)!}\sum_{N=2}^{T+1}N^{D-1}\frac{1}{(Dp)^{N}}
\end{aligned}\label{eq:n_a}
\end{equation}

To get rid of the summation in Eq.~\eqref{eq:n_a}, we approximate it by adding the negligible $N=1$ term, and replace the summation with an integral: 

\begin{equation}
\begin{aligned}
    \langle n_{\mathrm{a}, T}\rangle &\approx \frac{(1-p)^D (Dp)^{T+1}}{(D-1)!}\int_{0}^{T+1}x^{D-1}\frac{1}{(Dp)^{x}} \dd x\\
    &= \frac{(1-p)^D (Dp)^{T+1}}{(D-1)!}\int_{0}^{T+1}x^{D-1} e^{-\ln(Dp)x} \dd x\\
    &= \frac{(Dp)^{T+1}}{(D-1)!}\left(\frac{1-p}{\ln Dp}\right)^D\int_{0}^{(T+1)\ln Dp}x^{D-1} e^{-x} \dd x\\
    &= \frac{(Dp)^{T+1}}{(D-1)!}\left(\frac{1-p}{\ln Dp}\right)^D\gamma(D, (T+1)\ln Dp)
\end{aligned}\label{eq:n_a_1}
\end{equation}
where $\gamma(s, x)=\int_0^x t^{s-1}e^{-t}\dd t$ is the lower incomplete gamma function, and the result is valid for $Dp>1$. 

To further simplify $\langle n_{\mathrm{a}, T}\rangle$, we use the relation~\cite{jameson2016incomplete}
\begin{equation}
    \Gamma(n+1, x)\equiv\Gamma(n+1)-\gamma(n+1, x)=n! e_n(x) e^{-x}
\end{equation}
where $e_n(x)=\sum_{k=0}^n\frac{x^k}{k!}$ is the truncated exponential series.
Then, 
\begin{equation}
\begin{aligned}
    &\gamma(D, (T+1)\ln Dp) \\
    =&(D-1)!\left(1-\frac{e_D\big((T+1)\ln Dp\big)}{e^{(T+1)\ln Dp}}\right)=\alpha (D-1)!\,,    
\end{aligned}
\end{equation}
where
\begin{equation}
    \alpha=\left(1-\frac{e_D\big((T+1)\ln Dp\big)}{e^{(T+1)\ln Dp}}\right)\label{eq:alpha},
\end{equation}
is a number between 0 and 1.

Let us define 
\begin{equation}
    g(k, x)=\frac{x^k}{k!} e^{-x}.
\end{equation}

Using Stirling's formula, we have 
\begin{equation}
\begin{aligned}
    g(k, x)=&\frac{e^{-x}}{\sqrt{2\pi k}}\left(\frac{ex}{k}\right)^k\\
    =& \frac{1}{\sqrt{2\pi k}}e^{k(1+\ln x-\ln k)-x}.
\end{aligned}
\end{equation}

\begin{figure}
    \centering
    \includegraphics[width = 0.48\textwidth]{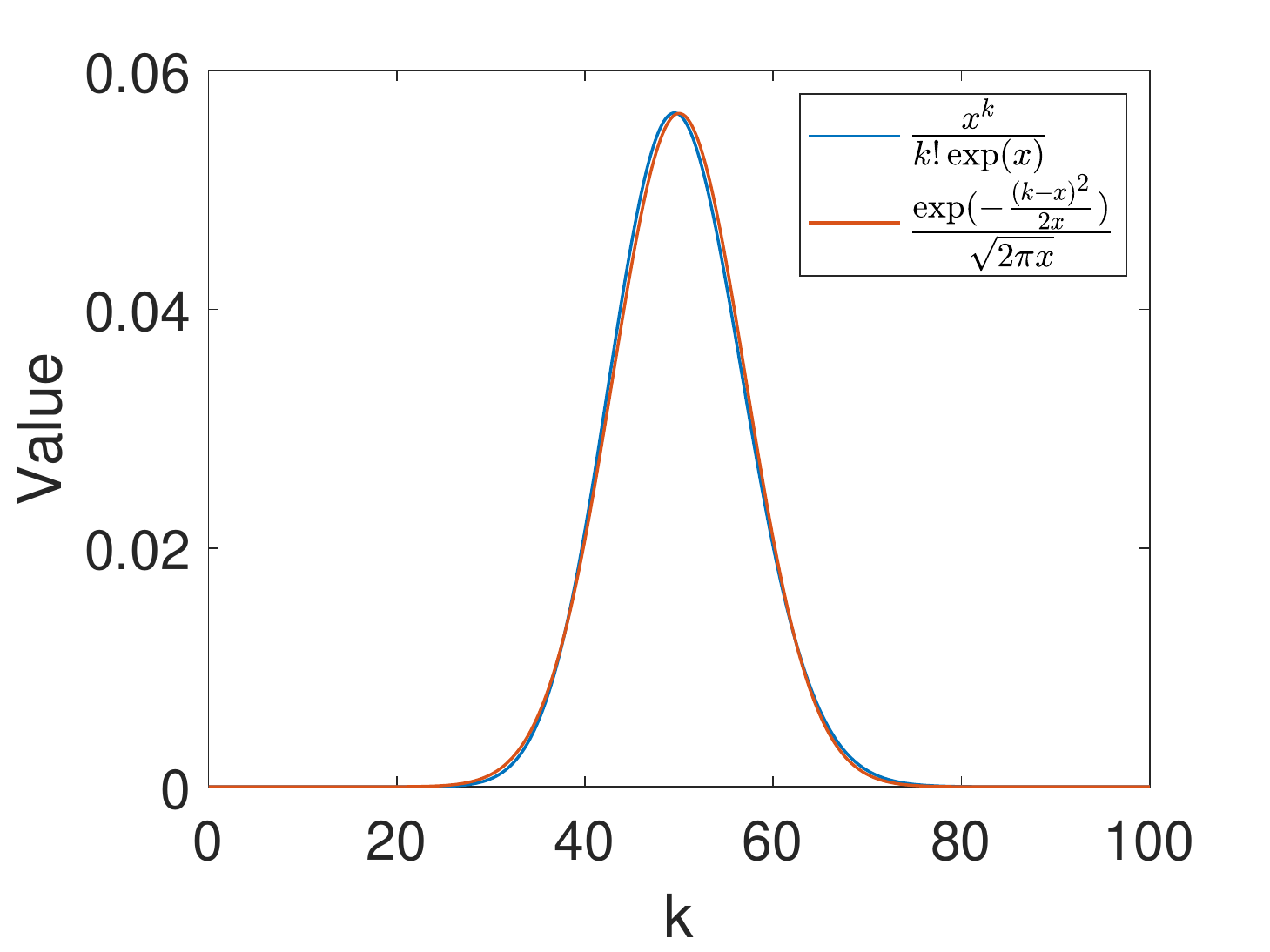}
    \caption{Comparison of each term in the truncated exponential series and the normal distribution. In this figure $x=50$.}
    \label{fig:gaussian}
\end{figure}

We can check that $g(k, x)$, as a function of $k$, reaches its maximum near $k=x$. Let us expand $\ln k$ near $k=x$:
\begin{equation}
    \ln k = \ln x + \frac{k-x}{x} -\frac{(k-x)^2}{2x^2} + O((\frac{k-x}{x})^3).
\end{equation}
Then, 
\begin{equation}
\begin{aligned}
    g(k, x)=&\frac{1}{\sqrt{2\pi k}}e^{k(1+\ln x-\ln x -\frac{k-x}{x}+\frac{(k-x)^2}{2x^2}+O((\frac{k-x}{x})^3))-x}\\
    =& \frac{1}{\sqrt{2\pi k}} e^{-k^2/x+2k-x+k(k-x)^2/(2x^2)+O((\frac{k-x}{x})^3)}\\
    =& \frac{1}{\sqrt{2\pi k}} e^{-\frac{(k-x)^2}{2x}(2-k/x+O(\frac{k-x}{x}))}.
\end{aligned}
\end{equation}
Near the maximum $k=x$, we have
\begin{equation}
    g(k, x)\approx\frac{1}{\sqrt{2\pi x}}e^{-(k-x)^2/2x},
\end{equation}
which is a normal distribution with mean value $x$ and standard deviation $\sqrt{x}$. Figure~\ref{fig:gaussian} shows the comparison of the original $g(k, x)$ and its approximation. We can see that the approximation is very good. 

Then, we have
\begin{equation}
\begin{aligned}
    \frac{e_n(x)}{e^x}= &\sum_{k=0}^n g(k, x)\\
    \approx & \int_0^n \frac{1}{\sqrt{2\pi x}}e^{-(k-x)^2/2x} \dd k\\
    \approx & \int_{-\infty}^n \frac{1}{\sqrt{2\pi x}}e^{-(k-x)^2/2x} \dd k\\
    = & \Phi(\frac{n-x}{\sqrt{x}}),
\end{aligned}
\end{equation}
where $\Phi(x)=\frac{1}{\sqrt{2\pi}}\int_{-\infty}^x e^{-t^2/2}\dd t$ is the cumulative distribution function of the standard normal distribution. 

Back to Eq.~\eqref{eq:alpha}, we finally have
\begin{equation}
\begin{aligned}
    \alpha =& 1-\frac{e_D\big((T+1)\ln Dp\big)}{e^{(T+1)\ln Dp}}\\
    =& 1-\Phi(\frac{D-(T+1)\ln Dp}{\sqrt{(T+1)\ln Dp}})\\
    =& \frac{1}{2}\erfc(\frac{D-(T+1)\ln Dp}{\sqrt{2(T+1)\ln Dp}}),
\end{aligned}\label{eq:alpha_1}
\end{equation}
where $\erfc$ is the complementary error function. 

In DMMs, each instanton connects two critical points whose indices (number of unstable directions) differ by 1 \cite{seanpaper}. Since the index of a critical point equals at most the dimension $D$, and the index of the equilibrium point is $0$, we have $T+1=D$, and Eq.~\eqref{eq:alpha_1} simplifies to
\begin{equation}
    \alpha =\frac{1}{2}\erfc(\sqrt{D}\frac{1-\ln Dp}{\sqrt{2\ln Dp}}).
\end{equation}

Therefore, 
\begin{equation}
    \langle n_{\mathrm{a}, T}\rangle=\frac{(Dp)^{T+1}}{2}\left(\frac{1-p}{\ln Dp}\right)^D\erfc(\sqrt{D}\frac{1-\ln Dp}{\sqrt{2\ln Dp}}),
\end{equation}
which is Eq.~(\ref{eq:na}) in the main text. 
\bibliographystyle{apsrev4-1}
\bibliography{SUSYref}

\end{document}